\documentclass[%
 reprint,
nofootinbib,
 amsmath,amssymb,
 aps,
]{revtex4-1}
\usepackage{graphicx}% Include figure files
\usepackage{dcolumn}% Align table columns on decimal point
\usepackage{bm}% bold math
\begin{document}

\preprint{APS/123-QED}

%\title{\boldmath Enhanced $B_d^0 \to \mu^+\mu^-$ Decay: What if?}

\title{\boldmath
Correlating $B_q^0 \to \mu^+\mu^-$ and $K_L \to \pi^0\nu\bar\nu$ Decays
with Four Generations}

\author{Wei-Shu Hou$^{a}$, Masaya Kohda$^{a,b}$, and Fanrong Xu$^{a,c}$} %,c}$}
 \affiliation{
$^{a}$Department of Physics, National Taiwan University,
 Taipei, Taiwan 10617\\
$^{b}$Department of Physics, Chung-Yuan Christian University,
 Chung-Li, Taiwan 32023\\
$^{c}$
Department of Physics, Jinan University, Guangzhou 510632, China
%Institute of Physics, Academia Sinica,
%Taipei, Taiwan 11529
%$^{c}$Department of Physics, Liaoning Normal University, Dalian 116029,
%P.R. China
}


\begin{abstract}
The $B_s\to \mu^+\mu^-$ mode has finally been observed,
albeit at rate 1.2$\sigma$ below Standard Model (SM) value,
while the rarer $B_d^0 \to \mu^+\mu^-$ decay has central value
close to 4 times SM expectation but with only 2.2$\sigma$ significance.
The measurement of $CP$ violating phase $\phi_s$
%, via both $B_s\to J/\psi K\bar K$ and $J/\psi\pi\pi$ modes,
has finally reached SM sensitivity.
Concurrent with improved measurements at LHC Run 2,
$K_L\to\pi^0\nu\bar\nu$ and $K^+\to\pi^+\nu\bar\nu$
decays are being pursued in a similar time frame.
We find, whether $B_d^0 \to \mu^+\mu^-$ is enhanced or not,
$K_L\to\pi^0\nu\bar\nu$ can be enhanced up to the Grossman-Nir bound
in the fourth generation model,
correlated with some suppression of $B_s\to \mu^+\mu^-$,
and with $\phi_s$ remaining small.
\begin{description}
% Secondary publications and information retrieval purposes.
\item[PACS numbers]
14.65.Jk % Other quarks (e.g., 4th generations)
12.15.Hh % Determination of Cabibbo-Kobayashi & Maskawa (CKM) matrix elements
11.30.Er % Charge conjugation, parity, time reversal, and other discrete symmetries
13.20.He % Decays of bottom mesons (Leptonic,semileptonic, and radiative decays of mesons)
% You may use the \texttt{description} environment to structure your abstract; use the optional argument of the \verb+\item+ command to give the category of each item.
\end{description}
\end{abstract}

\pacs{Valid PACS appear here}% PACS, the Physics and Astronomy
\maketitle

%\tableofcontents

\section{\label{sec:Intro}INTRODUCTION\protect\\}

The 7-and-8 TeV run (Run 1) of the LHC has been a great success,
but no New Physics (NP) has emerged.
The hint of NP in the forward-backward asymmetry
of $B\to K^*\ell^+\ell^-$ decay~\cite{Wei:2009zv} from the B factories
was eliminated by LHCb~\cite{Aaij:2011aa} early on. %with just 0.37 fb$^{-1}$ data.
The mild hint at the Tevatron~\cite{Aaltonen:2007-2012, Abazov:2008-2012}
for \emph{large} $CP$ violating (CPV) phase $\phi_s$ in $B_s^0$ mixing
was also swiftly eliminated %by the combined measurements of the
%$B_s^0\to J/\psi\,\phi$~\cite{LHCb:2011aa} and
%$B_s^0\to J/\psi\,f_0(980)$~\cite{LHCb:2011ab}
by LHCb~\cite{LHCb:2011aa,LHCb:2011ab},
% channels with %0.37 fb$^{-1}$ and 0.41 fb$^{-1}$ similar amount of LHCb data,
%, respectively.
vanquishing the suggested possible correlation~\cite{Hou:2006mx}
with large direct CPV difference $\Delta {\cal A}_{K\pi} \equiv
 {\cal A}(B^+\to K^+\pi^0) -  {\cal A}(B^0\to K^+\pi^-)$~\cite{Lin:2008zzaa}.
%\emph{if} the source for the latter arises from the electroweak penguin~\cite{Deshpande:1998xq},
%was vanquished.
%
Finally, the hot pursuit for $B_s^0\to \mu^+\mu^-$ at the Tevatron
culminated in the recent observation by the LHCb~\cite{Aaij:2013aka}
and CMS~\cite{Chatrchyan:2013bka} experiments, %~\cite{foot1},
albeit again consistent with the Standard Model (SM).
%This put an end to the speculation from minimal supersymmetry
%that the rate could have been enhanced by three orders of magnitude.

The combined LHC result for $B_q^0\to \mu^+\mu^-$
is~\cite{B2mumu-LHCbCMS},
\begin{align}
{\cal B}(B_s^0\to \mu^+\mu^-) &= (2.8^{+0.7}_{-0.6}) \times 10^{-9},
 \label{eq:BsmumuLHC} \\
{\cal B}(B_d^0\to \mu^+\mu^-) &= (3.9^{+1.6}_{-1.4}) \times 10^{-10}.
 \label{eq:BdmumuLHC}
\end{align}
At 6.2$\sigma$, the $B_s^0\to \mu^+\mu^-$ mode is established,
but SM expectation is 7.6$\sigma$.
The $B_d^0\to \mu^+\mu^-$ mode %is nonzero at the 3.2$\sigma$ level, but
deviates from SM expectation of $(1.06 \pm0.09) \times 10^{-10}$~\cite{Bobeth:2013uxa}
by 2.2$\sigma$,
%While this should not be taken too seriously,
with central value more than $3$ times the SM value.
Thus, $B_d^0\to \mu^+\mu^-$ should be keenly followed at the
up and coming LHC Run 2 (13 and 14 TeV).

The 1 fb$^{-1}$ LHCb update for $\phi_s$~\cite{Aaij:2013oba} is:
\begin{equation}
\phi_s = 0.01 \pm 0.07 \pm 0.01,\ \ \ {\rm (1\ fb}^{-1},\ {\rm LHCb)}
 \label{eq:phis_LHCb-1}
\end{equation}
which combines two results opposite in sign,
%both the $B_s^0\to J/\psi\,\phi$ and $J/\psi\,\pi^+\pi^-$ channels, with respective values
%
\begin{align}
\phi_s &= 0.07 \pm 0.09 \pm 0.01,\ \; {\rm (1\ fb}^{-1}\ J/\psi K\bar K,\ {\rm LHCb)}
 \label{eq:phis_LHCb-1-KK} \\
\phi_s &= -0.14^{+0.17}_{-0.16} \pm 0.01.\ \ {\rm (1\ fb}^{-1}\ J/\psi\,\pi\pi,\ {\rm LHCb)}
 \label{eq:phis_LHCb-1-pipi}
\end{align}
%
%which are of opposite sign.
%Eq.~(\ref{eq:phis_LHCb-1-pipi})
%is improved from an earlier~\cite{LHCb:2012ad} 1~fb$^{-1}$ result
%of $-0.019^{+0.173+0.004}_{-0.174-0.003}$,
%mainly due to improved tagging.
% of the $B_s^0$ flavor, i.e. $B_s^0$ vs $\overline B_s^0$.
%With same tagging, Eq.~(\ref{eq:phis_LHCb-1-KK}) is improved from
%the 0.37 fb$^{-1}$ result of $0.15\pm0.18\pm0.06$.
%
Eq.~(\ref{eq:phis_LHCb-1}) dominates the Heavy Flavor Averaging Group (HFAG)
combination~\cite{Amhis:2012bh} of all experiments,
\begin{equation}
\phi_s = 0.00 \pm 0.07,\ \ \ {\rm (PDG2014)}
 \label{eq:phis_PDG2014}
\end{equation}
adopted by the Particle Data Group (PDG)~\cite{PDG14},
and is in good agreement with SM value of $\phi_s \simeq -0.04$.
A preliminary result~\cite{CMS:2014jxa} of CMS, %based on 20 fb$^{-1}$ data,
\begin{equation}
\phi_s = -0.03 \pm 0.11 \pm 0.03,\ \ {\rm (20\ fb}^{-1},\ {\rm CMS)}
 \label{eq:phis_CMS-20}
\end{equation}
is not included in PDG2014, but is fully consistent.

Also not making PDG2014 is the 3 fb$^{-1}$ update by LHCb for
$B_s^0\to J/\psi\,\pi^+\pi^-$ with full Run 1 data~\cite{Aaij:2014dka},
\begin{equation}
\phi_s = 0.070 \pm 0.068 \pm 0.008,\ \ {\rm (3\ fb}^{-1}\ J/\psi\,\pi\pi,\ {\rm LHCb)}
 \label{eq:phis_LHCb-3-pipi}
\end{equation}
%
%This is the single best measurement so far, but
with sign change from Eq.~(\ref{eq:phis_LHCb-1-pipi}),
%one sees that adding twice more data
%lead to not only a large reduction of error,
%but a change in sign of the central value,
becoming same sign with Eq.~(\ref{eq:phis_LHCb-1-KK}).
%This is again in part due to improved tagging.
Because the analysis is done simultaneously
with $\Delta\Gamma_s$ measurement,
the $B_s^0\to J/\psi\,K\bar K$ mode took much longer.
%If the 3 fb$^{-1}$ update for $B_s^0\to J/\psi\,\phi$
%keeps the same sign as Eq.~(\ref{eq:phis_LHCb-1-KK}),
%then $\phi_s$ would become positive in sign, and would
%start to deviate obviously from SM expectation!
%
%and the positive tendency of world average was given
%in an unofficial combination at a conference~\cite{phi_s-BEACH}.
%
%In this vein, it is good that the CMS as well as ATLAS experiments,
%even though not quite competitive with LHCb (see Eq.~(\ref{eq:phis_CMS-20}))
%on $\phi_s$, would provide a sanity check.
%
Intriguingly, it also switched sign~\cite{LHCb_1411},
\begin{equation}
\phi_s = -0.058 \pm 0.049 \pm 0.006,\ \; {\rm (3\ fb}^{-1}\; J/\psi\, \phi,\; {\rm LHCb)}
 \label{eq:phis_LHCb-3-KK}
\end{equation}
%
%(cf. Eq.~(\ref{eq:phis_LHCb-1-KK})),
and the combined 3~fb$^{-1}$ result is,
\begin{equation}
\phi_s = -0.010 \pm 0.039, \ \ \ \rm{(3~fb}^{-1},\ {\rm LHCb})
 \label{eq:phis_LHCb-3}
\end{equation}
with no indication of New Physics.
%
%With the $B_s^0 \to J/\psi\phi$ and $B_s^0 \to J/\psi\pi\pi$
%mode both fluctuating in sign as data increases,
%it must be that the $\phi_s$ value is very tiny,
%as reflected in the central value of Eq.~(\ref{eq:phis_LHCb-3}),
%rather than what was indicated in Ref.~\cite{phi_s-ICHEP},
%or Eq.~(\ref{eq:LHCb-ICHEP2014}).
%There is no hint of New Physics,
%but it also does not change the results and
%presentation of our paper.
%In some sense, it is a ``vindication'' of our
%caution that $\phi_s > 0$ is not particularly plausible.

% given that the central value of
%Eq.~(\ref{eq:BdmumuLHC}) for $B_d^0\to \mu^+\mu^-$ is
%more than 3 times the SM expectation,
%of $(1.06 \pm0.09) \times 10^{-10}$~\cite{Bobeth:2013uxa},
%it is imperative to follow up with Run 2 data,
%and the experiments should hone their
%analyses on $B_d^0\to \mu^+\mu^-$, now that the $B_s^0\to \mu^+\mu^-$
%mode is more or less ``done'' (although it should still be watched).
%; the mild suggestion of suppression of
%the latter would take considerably more data to elucidate.
%
A third measurement of interest by LHCb is
the so-called $P_5^\prime$ anomaly.
The significance (3.7$\sigma$), however, did not change
from 1 fb$^{-1}$~\cite{Aaij:2013qta} to 3 fb$^{-1}$~\cite{P5'-Moriond}.
%, which is a bit worrisome.
Given that this is one out of many angular variables in $B^0\to K^{*0}\mu^+\mu^-$,
it remains to be seen whether the ``anomaly'' is genuine.
Unfortunately, it will take several years to accumulate and analyze
an equivalent amount of data at Run~2.
We note in passing the so-called $R_K$ anomaly~\cite{Aaij:2014ora}
in lepton universality violation in $B^+\to K^+\ell^+\ell^-$ decays,
which has 2.6$\sigma$ deviation from SM expectation.
It may or may not be related to the $P_5^\prime$ anomaly.

%Whether $P_5^\prime$ or $R_K$ stay interesting as Run 2 data unfolds
%is unclear at present. But it should be clear that
%$B_q^0\to \mu^+\mu^-$ and $\phi_s$ measurements would remain
%lampposts for New Physics.
%
What are the prospects for Run 2?
A total of $8$ fb$^{-1}$ or more data is expected by LHCb up to 2018.
Data rate is much higher for CMS, but trigger bandwidth
is an issue.
%, although things would certainly be extremely interesting.
%
The Belle~II experiment should be completed in this time frame in Japan.
Though not particularly competitive in $\phi_s$ and $B_q^0\to \mu^+\mu^-$,
it could crosscheck $P_5^\prime$.
Given that the two former measurables correspond to
$b \leftrightarrow s$ and $b\to d$ transitions,
one involving CPV, the other not,
there is one particular process that comes to mind:
$K\to \pi\nu\bar\nu$ decays, which are $s\to d$ transitions.
The neutral $K_L^0 \to \pi^0\nu\nu$ decay,
pursued by the KOTO experiment~\cite{KOTO} in Japan, is purely CPV.
The charged $K^+\to \pi^+\nu\nu$ mode is pursued
by the NA62 experiment~\cite{NA62} at CERN.
Both experiments run within a similar time frame.
If one has indications for NP in $B_q^0\to \mu^+\mu^-$ and/or $\phi_s$,
likely one would find NP in $K\to \pi\nu\bar\nu$, and \emph{vice versa}.
An element of competition between high- and low-energy
luminosity frontiers would be quite interesting.

In this letter we study the \emph{correlations} between
the measurables $B_d^0\to \mu^+\mu^-$, $B_s^0\to \mu^+\mu^-$,
$\phi_s$, and %(CPV phase in $B_s$ mixing)
$K\to \pi\nu\bar\nu$ (especially $K_L^0 \to \pi^0\nu\nu$),
in the 4th generation (4G) model, which cannot address $P′_5/R_K$ anomalies.
It was pointed out quite some time ago~\cite{Hou:2005yb} that
4G can bring about an enhanced $K_L^0 \to \pi^0\nu\nu$,
and now that KOTO is running, one should check
whether it remains true.
Although some may now find 4G extreme, our aim is towards
enhanced $B_d^0\to \mu^+\mu^-$ rate by a factor of three
and still survive all \emph{flavor} constraints.
The issue with 4G is the observation of a light Higgs boson,
without the anticipated factor of 9 enhancement in cross section.
On one hand it has been argued~\cite{MHK} that  there 
 still exists other interpretation of this 125 GeV boson, 
that is to identify it as dilaton from a  4G theory with strong Yukawa interaction.
On the other hand, 
Higgs boson practically does not enter (i.e. is ``orthogonal" to)
low energy flavor changing processes, and, \emph{if}
one discovers an enhanced $B_d^0\to \mu^+\mu^-$ decay~\cite{Hou:2013btm},
it may put some doubt on the Higgs nature of the observed
125 GeV particle. We view 
the issue, different interpretation of this boson, is
still opens and would 
 be settled by 2018.
Our 4G study serves to illustrate how New Physics in
$B_q^0\to \mu^+\mu^-$, $\phi_s$, and $K\to \pi\nu\bar\nu$
might be accommodated.

%The paper is organized as follows. We review
In what follows, we give %in Sec.~II
the formulas and data inputs,
then our numerical results %given in Sec.~III.
%Given that $\phi_s$ value for full Run 1 data is still
%quite volatile, we take the more conservative approach
%of Eq.~(\ref{eq:phis_PDG2014}),
%but do explore the situation for $\phi_s > 0$.
and end with some discussions. %and give our conclusion in Sec.~IV.

\section{Formulas and Data Input}

%Unless stated otherwise, we use experimental values from
%PDG 2013 partial update~\cite{PDG13}.
%As our purpose is to illustrate correlation between
%$B_q^0 \to \mu^+\mu^-$ and $K \to \pi\nu\bar\nu$,
%updating to PDG2014 does not make major difference.
%
We define the parameters
$x_q = {m_q^2}/{M_W^2}$,
$\lambda_q^{ds}\equiv V_{qd}V_{qs}^*$ ($q=u,c,t,t^\prime$),
with
\begin{align}
%x_q = \frac{m_q^2}{M_W^2}, \quad &
%&\lambda_q^{ds}\equiv V_{qd}V_{qs}^* \quad (q=u,c,t,t^\prime), \\
%\end{equation}
%\begin{equation}
V_{t^\prime d}^*V_{t^\prime s} & \equiv (\lambda_{t^\prime}^{ds})^*
\equiv r_{ds}e^{i\phi_{ds}}.
 \label{eq:rdsphids}
\end{align}
We adopt the parametrization of Ref.~\cite{Hou:1987hm} for
the $4\times 4$ CKM matrix, with convention and treatment of Ref.~\cite{Hou:2005yb}. In particular,
we assume SM-like values for $s_{12}$, $s_{23}$, $s_{13}$
and $\phi_{ub}\simeq \gamma/\phi_3$,
with following input: %~\cite{PDG13}
%\begin{align}
$|V_{us}|=0.2252\pm 0.0009$, %\ \ \ \
$|V_{cb}|= 0.0409 \pm 0.0011$, %\notag \\
$|V_{ub}^{\rm ave}|=(4.15\pm 0.49)\times 10^{-3}$, %\ \
$\gamma/\phi_3 = (68^{+10}_{-11})^\circ$.
% \notag
%\end{align}
This is a simplification, since we try to observe trends,
rather than making a full fit.
We find taking the ``exclusive'' measurement value %~\cite{PDG13}
for $|V_{ub}|$ %would be over-exclusive, with no meaningful solution space.
allows less enhancement range for $K_L\to \pi^0\nu\bar\nu$.

Having a 4th generation of quarks brings in
three new angles and two new phases.
In this paper, we take
\begin{equation}
m_{t^\prime} = 1000~{\rm GeV}, \quad
s_{34} \simeq {m_W}/{m_{t^\prime}} \simeq 0.08,
 \label{eq:tp_s34}
\end{equation}
for sake of illustration, thereby fixing one of the angles.
A second angle and one of the two phases are
fixed by the discussion illustrated below.
We are then left with two mixing parameters,
and for our interest in $K\to \pi\nu\bar\nu$ decays,
we take as $r_{ds}$ and $\phi_{ds}$ in Eq.~(\ref{eq:rdsphids}).
%\\
%
Our choice~\cite{foot} of $m_{t^\prime} = 1000$~GeV
is above the experimental bound~\cite{PDG14},
which is beyond the nominal~\cite{Chanowitz:1978uj} unitarity bound.
%Experiments are now vigorously exploring the so-called
%vector-like quarks, with signatures beyond $b'\to tW$ and $t'\to bW$
%(involving tree level FCNC %flavor changing neutral current
%decays to $Z$ and Higgs bosons).
Even with vector-like 4G quarks,
the experimental bound has reached beyond 700~GeV~\cite{PDG14}.
We adhere to sequential 4G to reduce the number of parameters.

\begin{figure}[t!]
%\begin{center}
%\includegraphics[width=60mm,height=40mm]{plots/twidth.pdf}
%\vspace{30mm}
{
 \includegraphics[width=70mm]{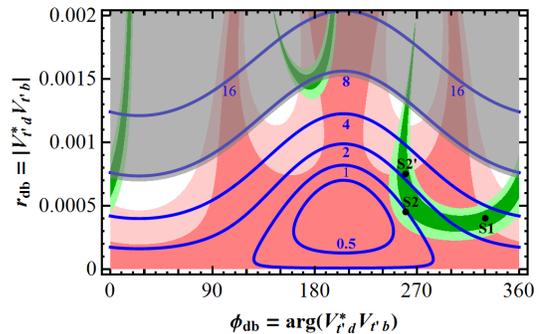}
}
%\end{center}
\vskip0.55cm
\caption{
Update of Fig.~3(a) of Ref.~\cite{Hou:2013btm},
taking $|V_{ub}^{\rm ave}|$ and $m_{t'} = 1000$ GeV.
The pink-shaded contours correspond to
1(2)$\sigma$ regions of $\Delta m_{B_d}$ allowed by
$f_{B_d} = (190.5\pm 4.2)$ MeV %Eq.~(\ref{eq:fBd}),
while the green-shaded bands are for
1(2)$\sigma$ in $\sin2\beta/\phi_1 = 0.682 \pm 0.019$~\cite{Amhis:2012bh}. %(Eq.~(\ref{eq:sin2beta})).
Solid-blue lines are labeled $10^{10}\mathcal B(B_d\to \mu^+\mu^-)$
contours, with upper bound of 7.4~\cite{Aaij:2013aka} applied.
Marked points S1, S2, S2$^\prime$ are explained in text.
} \label{fig:b2d}
\end{figure}
%

%
%\noindent
%\underline{Update of Fig. 3 in HKX 2013}
%\\

We have suggested~\cite{Hou:2013btm} that an enhanced
$B_d\to\mu^+\mu^-$ could indicate the presence of 4G,
while suppression of $B_s \to \mu^+\mu^-$
could also be accommodated~\cite{Hou:2012xe}.
These are supported by current data~\cite{B2mumu-LHCbCMS},
but more data is clearly needed.
In Fig.~\ref{fig:b2d}, we update Fig.~3(a) of Ref.~\cite{Hou:2013btm}
on the $r_{db}$--$\phi_{db}$ plane, where
$V_{t^\prime d}^*V_{t^\prime b}\equiv r_{db}e^{i\phi_{db}}$.
We use the FLAG~\cite{Aoki:2013ldr} average of $N_f =2+1$ lattice results
%\begin{align}
% f_{B_d} &= (190.5\pm 4.2)~{\rm MeV},  \\
% f_{B_d}\sqrt{\hat B_{B_d}} &= (216\pm 15)~{\rm MeV},
% \label{eq:fBd}
%\end{align}
%for hadronic parameters.
$f_{B_d} = (190.5\pm 4.2)$ MeV,
$f_{B_d}{\hat B_{B_d}}^{1/2} = (216\pm 15)$ MeV.
We no longer take the ratio with $\Delta m_{B_d}$
for $B_d\to\mu^+\mu^-$ branching ratios,
but update the constraints:
%\begin{align} &
$\sin2\beta/\phi_1 = 0.682\pm 0.019$ (HFAG~Winter~2014~\cite{Amhis:2012bh}),
 %\label{eq:sin2beta} \\
%&
$\mathcal B(B_d\to\mu^+\mu^-) < 7.4\times 10^{-10}$ (95\% CL~limit of LHCb~\cite{Aaij:2013aka}).
% \label{eq:upBdmumu}
%\end{align}
%and, respectively.
The latter is softer than the recent CMS and LHCb combination
of Eq.~(\ref{eq:BdmumuLHC}).

From Fig.~\ref{fig:b2d}, we shall consider two scenarios
\begin{equation}
r_{db}e^{i\phi_{db}}=0.00040 e^{i\,330^\circ},
 \ 0.00045 e^{i\,260^\circ},
% \ 0.00075 e^{i\,260^\circ}
 \label{eq:rdbphidb}
\end{equation}
marked as S1, S2 %and S2$^\prime$
in Fig.~\ref{fig:b2d}, to illustrate
\begin{equation}
 \mathcal B(B_d\to \mu^+\mu^-)
  \sim 3\times 10^{-10}, \ \;
       1\times 10^{-10}, \ \;
%       2\times 10^{-10},
\end{equation}
where we stay within
1$\sigma$ boundaries of \emph{both} $\Delta m_{B_d}$
(uncertainty in $f_{B_d}$) %Eq.~(\ref{eq:fBd}))
and $\sin2\beta/\phi_1$.
$B_d\to \mu^+\mu^-$ is SM-like for S2,
but carries a near maximal 4G CPV phase $\phi_{db}$.
The point S2$^\prime$
%can reach $\mathcal B(B_d\to \mu^+\mu^-) \sim 2\times 10^{-10}$, by
%having a real part for $e^{i\,\phi_{db}}$, or
 %for larger $r_{db}$
 will be discussed towards the end.

For $b\to s$ observables, we update both formulas
and input parameters of Ref.~\cite{Hou:2011fw}.
For the CPV phase $\phi_s\equiv 2\Phi_{B_s}$ in $B_s$-$\bar B_s$ mixing,
we use~\cite{Buras:2010pi} $2\Phi_{B_s} = \arg \Delta_{12}^s$
with
%\begin{align}
%2\Phi_{B_s} &= \arg \Delta_{12}^s, \\
$\Delta_{12}^s
%&
= (\lambda_t^{bs})^2 S_0(x_t)
  +2\lambda_t^{bs}\lambda_{t^\prime}^{bs}S_0(x_t,x_{t^\prime}) %\notag %\\
 %& \hskip1.5cm
 +(\lambda_{t^\prime}^{bs})^2 S_0(x_{t^\prime})$, %\notag
%\end{align}
where $\lambda_q^{bs} \equiv V_{qs}^*V_{qb}$ ($q=t,t^\prime$).
We adopt the $\phi_s$ value of Eq.~(\ref{eq:phis_LHCb-3})
and impose $1(2)\sigma$ constraints.

For $\mathcal B(B_s\to \mu^+\mu^-)$,
because $\Delta\Gamma_s$ is sizable,
experimental results should be compared with the
$\Delta\Gamma_s$-corrected~\cite{DeBruyn2012}
branching ratio denoted with a bar,
%which is related to uncorrected one without bar via~\cite{Buras:2013ooa}
\begin{equation}
\bar{\mathcal B}(B_s\to \mu^+\mu^-)
= %\frac{1}{r(y_s)}
\frac{1-y_s^2}{1+ \mathcal A_{\Delta \Gamma}^{\mu\mu}y_s} \mathcal B(B_s\to \mu^+\mu^-),
\end{equation}
where $y_s = \Delta\Gamma_s/2\Gamma_s=0.069\pm 0.006$ \cite{PDG14},
%\begin{align}
%r(y_s)&\equiv \frac{1-y_s^2}{1+ \mathcal A_{\Delta \Gamma}^{\mu\mu}y_s}, \\
$A_{\Delta \Gamma}^{\mu\mu} %&
= \,\cos[2\arg(C_{10}) -2\Phi_{B_s}]$,
%\end{align}
and~\cite{Buras:2010pi}
\begin{align}
&\mathcal B (B_s\to \mu^+\mu^-)
= \tau_{B_s} \frac{G_F^2}{\pi}\left( \frac{\alpha}{4\pi\sin^2\theta_W} \right)^2
f_{B_s}^2m_\mu^2m_{B_s} \notag \\
& \hskip0.15cm \times\sqrt{1-{4m_\mu^2}/{m_{B_s}^2}}\,\eta_{\rm eff}^2
\left|\lambda_t^{bs}Y_0(x_t)+ \lambda_{t^\prime}^{bs}Y_0(x_{t^\prime})\right|^2.
\end{align}
We use $\eta_{\rm eff}=0.9882\pm 0.0024$
which is at NNLO for QCD and NLO for electroweak corrections~\cite{Buras:2013dea},
and we use the FLAG~\cite{Aoki:2013ldr} average of $N_f =2+1$ lattice results
%\begin{align}
% f_{B_s} &= (227.7\pm 4.5)~{\rm MeV},  \\
% f_{B_s}\sqrt{\hat B_{B_s}} &= (266\pm 18)~{\rm MeV},
% \label{eq:fBs}
%\end{align}
$f_{B_s} = (227.7\pm 4.5)$ MeV and
$f_{B_s}{\hat B_{B_s}^{1/2}} = (266\pm 18)$ MeV,
where the latter enters $B_s$-mixing.
We use %the combined~\cite{B2mumu-LHCbCMS} Run 1 result of
%CMS and LHCb is
%$\bar{\mathcal B}(B_s\to \mu^+\mu^-)_{\rm exp}
%=(2.8^{+0.7}_{-0.6})\times 10^{-9}$ is
%given in
Eq.~(\ref{eq:BsmumuLHC}) %, although without the bar,
and impose $1(2)\sigma$ experimental constraints,
which is much larger than the hadronic uncertainty.

We find that $\Delta m_{B_s}$ does not give further constraints
in the parameter space of our interest,
within hadronic uncertainty of $f_{B_s}$. %from Eq.~(\ref{eq:fBs}).
The ratio $\Delta m_{B_s}/\Delta m_{B_d}$ has reduced hadronic uncertainty,
as can be read from
$\xi \equiv f_{B_s}{\hat B_{B_s}^{1/2}}/f_{B_d}{\hat B_{B_d}^{1/2}}
= 1.268 \pm 0.063$~\cite{Aoki:2013ldr},
hence provides stronger constraint
than individual $\Delta m_{B_s}$ or $\Delta m_{B_d}$.
%due to.
%
%we find that our parameter space of interest is allowed
%within uncertainty from $\xi$ at $2\sigma$ level.

%
For $K^+\to\pi^+\nu\bar\nu$ and $K_L\to\pi^0\nu\bar\nu$,
we use the formulas of Ref.~\cite{Hou:2005yb}
and update input parameters:
$m_t(m_t)=163$ GeV~\cite{Hoang:2008yj},
$\kappa_+ = (5.173\pm 0.025) \times 10^{-11}\times(|V_{us}|/0.225)^8$ and
$\kappa_L = (2.231\pm 0.013) \times 10^{-10} \times(|V_{us}|/0.225 )^8$~\cite{Mescia:2007kn},
and $P_c(X) = 0.41\pm 0.05$~\cite{Buras:2004uu}.
We impose
%\begin{equation}
$\mathcal B(K^+\to\pi^+\nu\bar\nu)_{\rm exp}
< 3.35\times 10^{-10}$ (90\% C.L. from E949~\cite{Artamonov:2009sz}),
%\end{equation}
implying the Grossman-Nir (GN) bound~\cite{Grossman:1997sk}
\begin{align}
\mathcal B (K_L\to\pi^0\nu\bar\nu)
&
< \frac{\kappa_L}{\kappa_+}\mathcal B (K^+\to\pi^+\nu\bar\nu) \nonumber \\
%&\simeq 4.3\times\mathcal B (K^+\to\pi^+\nu\bar\nu)  \\
&\lesssim 1.4 \times 10^{-9}, %\ \ \ {\rm (GN\ bound)}
 \label{eq:GNbound}
\end{align}
%For $K_L\to\pi^0\nu\bar\nu$,
which is stronger than the direct limit
by E391a~\cite{Ahn:2009gb}. %, which is at, $2.6\times 10^{-8}$.

For Short-Distance (SD) contribution to $K_L\to \mu^+\mu^-$,
\begin{align}
&\mathcal B(K_L\to \mu^+\mu^-)_{\rm SD}
=\kappa_\mu|V_{us}|^{-10} \left[
 {\rm Re} \lambda_c^{ds}\, |V_{us}|^{4} P_c(Y) \right. \notag \\
 &\quad\quad\quad\quad \left. +\;{\rm Re} \lambda_t^{ds}\, \eta_Y Y_0(x_t)
 +{\rm Re} \lambda_{t^\prime}^{ds}\eta_Y Y_0(x_{t^\prime})
\right]^2,
\end{align}
we use $\kappa_\mu = (2.009\pm 0.017) \times 10^{-9}\times(|V_{us}|/0.225)^8$,
$P_c(Y)= (0.115\pm 0.018)\times(0.225/|V_{us}|)^4$~\cite{Gorbahn:2006bm},
and $Y_0(x)$ as given in Ref.~\cite{Buras:2010pi}.
With the common QCD correction factor $\eta_Y=1.012$,
we adopt the estimate~\cite{Isidori:2003ts}
%\begin{equation}
$\mathcal B(K_L\to \mu^+\mu^-)_{\rm SD}
\leq 2.5\times 10^{-9}$
%\end{equation}
as upper bound. %on the SD contribution.

For $K\to\pi\pi$ indirect CP violation,
we use~\cite{Buras:2010pi}.
\begin{align}
\varepsilon_K
=&\frac{\kappa_\varepsilon e^{i\varphi_\varepsilon}}{\sqrt{2}(\Delta m_K)_{\rm exp}}
 {\rm Im}(M_{12}^K), \\
(M_{12}^K)^*
=&\frac{G_F^2M_W^2}{12\pi^2}m_Kf_K^2\hat B_K \left[
 (\lambda_c^{ds})^2 \eta_{cc}S_0(x_c) \right. \notag\\
 &+(\lambda_t^{ds})^2 \eta_{tt}S_0(x_t)
 +2\lambda_c^{ds}\lambda_t^{ds} \eta_{ct}S_0(x_c,x_t) \notag\\
 & +(\lambda_{t^\prime}^{ds})^2 \eta_{t^\prime t^\prime}S_0(x_{t^\prime})
 +2\lambda_c^{ds}\lambda_{t^\prime}^{ds} \eta_{ct^\prime}S_0(x_c,x_{t^\prime}) \notag\\
 &\left. +\;2\lambda_t^{ds}\lambda_{t^\prime}^{ds} \eta_{tt^\prime}S_0(x_t,x_{t^\prime})
 \right],
\end{align}
where $(\Delta m_K)_{\rm exp}=(5.293\pm 0.009)\times 10^{9}~{\rm s}^{-1}$~\cite{PDG14},
$\varphi_{\varepsilon}= (43.52\pm 0.05)^\circ$~\cite{PDG14} and
$\kappa_{\varepsilon}=0.94\pm 0.02$~\cite{Buras:2010pza}.
%Definition for the loop functions $S_0(x)$ and $S_0(x,y)$ can be found in Ref. \cite{Buras:2010pi}.
We use %the QCD correction factors
$\eta_{cc}=1.87\pm 0.76$~\cite{Brod:2011ty},
$\eta_{tt}= 0.5765\pm 0.0065$~\cite{Buras:1990fn,Brod:2010mj},
$\eta_{ct}=0.496\pm 0.047$~\cite{Brod:2010mj}
and approximate $\eta_{tt}=\eta_{tt^\prime}=\eta_{t^\prime t^\prime}$,
$\eta_{ct}=\eta_{ct^\prime}$.
Theoretical uncertainty %for the SM prediction
is around $11$\%~\cite{Buras:2013ooa},
far larger than experimental error~\cite{PDG14}:
%\begin{equation}
$|\varepsilon_K| =(2.228\pm 0.011)\times 10^{-3}$ %\quad
and ${\rm Re}(\varepsilon_K)>0$.
%\end{equation}
We thus impose $\varepsilon_K$ to be within $\pm 11$\%
from data.

\begin{figure*}[t!]
\centering
%\includegraphics[width=60mm,height=40mm]{plots/twidth.pdf}
%\vspace{30mm}
{\includegraphics[width=80mm]{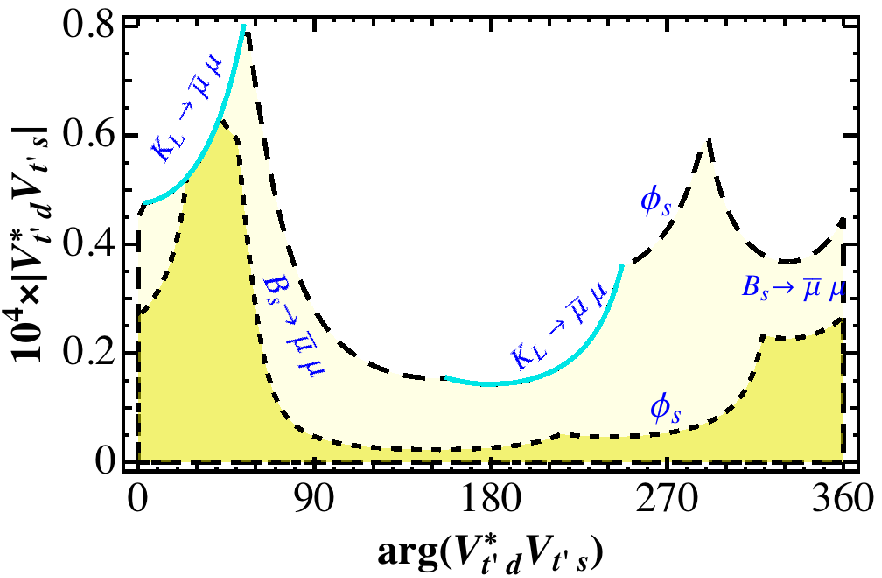}
 \includegraphics[width=80mm]{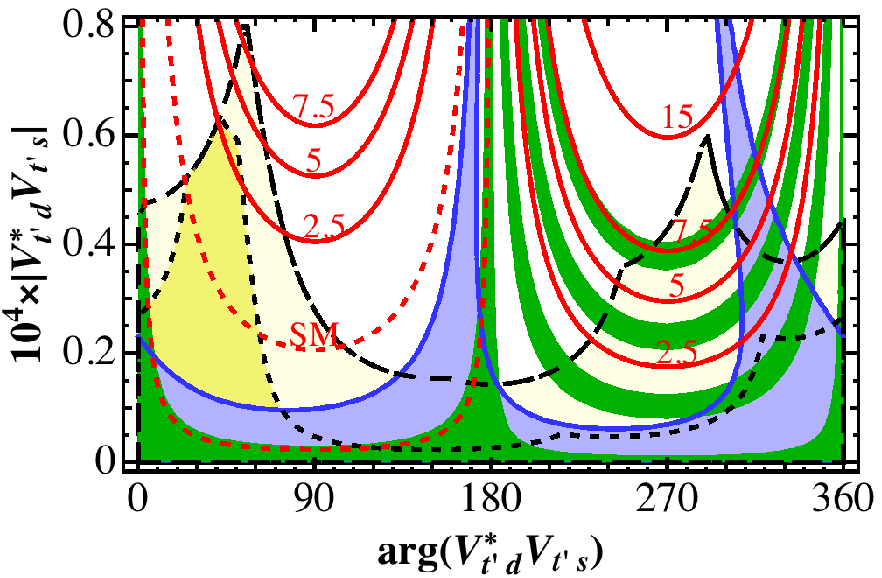}
}
\caption{
[left]
 Allowed region in $|V_{t'd}^*V_{t's}|$--$\arg (V_{t'd}^*V_{t's})$
 (i.e. $r_{ds}$--$\phi_{ds}$) plane for Scenario S1,
 $r_{db}\,e^{i\phi_{db}} = 0.0004\,e^{i330^\circ}$ (enhanced $B_d \to \mu^+\mu^-$)
   and $\phi_s = -0.010\pm 0.039$ (Eq.~(\ref{eq:phis_LHCb-3})),
 where the constraint source for each boundary is indicated.
 The leading constraint is $B_s \to \mu^+\mu^-$, where
 1(2)$\sigma$ region
 --- towards larger (smaller) BR in
  central region (4th-extending-to-1st quadrants) ---
 is (very) light shaded, separated by dashed lines, except:
  $K_L \to \mu^+\mu^-$ cuts off at upper left, as well as center-right,
  indicated by light-blue solid lines;
  1(2)$\sigma$ allowed $\phi_s$ cuts off
  the 1(2)$\sigma$ allowed $B_s \to \mu^+\mu^-$ in right-center,
  plus a sliver in 1st quadrant.
  %, while the $K^+\to \pi^+\nu\nu$ upper bound (solid purple line)
  %cuts off the 2$\sigma$ allowed $B_s \to \mu^+\mu^-$.
[right]
 The allowed region is further overlaid with $\varepsilon_K$ (blue-shaded),
 $\varepsilon'/\varepsilon$ (narrow green bands corresponding to
 $R_6$ in increasing order from 1.0, 1.5, 2.0, 2.5)
 and ${\cal B}(K_L\to \pi^0\nu\nu)$, labeled in $10^{-10}$ units.
 The illustration is for $m_{t'}= 1000$ GeV (Eq.~(11)).
 %, and other parameter choices are discussed in Sec.~II.
} \label{fig:fig2}
\end{figure*}

Direct CP violation in $K\to\pi\pi$ is affected
even more by hadronic uncertainties.
We use~\cite{Hou:2005yb}
\begin{align}
\frac{\varepsilon^\prime}{\varepsilon}
=a\left[ {\rm Im}(\lambda_c^{ds})P_0 +{\rm Im}(\lambda_t^{ds})F(x_t)
 +{\rm Im}(\lambda_{t^\prime}^{ds})F(x_{t^\prime}) \right],
\notag %\label{eq:DCPV4}
\end{align}
where $a=0.92\pm 0.03$~\cite{Buras:2014sba}
is a correction from $\Delta I=5/2$ transitions~\cite{Cirigliano:2003nn}.
The function $F(x)$, which relies on hadronic parameters $R_6$ and $R_8$\footnote{
Furthermore, the relations between hadronic parameters $R_{6,8}$ and bag parameters $B_{6}^{(1/2)}$ and
$B_8^{(3/2)}$ which are 
calculated in Lattice QCD, are $R_{6}=1.13B_{6}^{(1/2)} \left[\frac{114\, \text{MeV}}
{m_s(m_c)+m_d(m_c)}\right]^2$
and $R_8=1.13B_8^{(3/2)}\left[\frac{114\, \text{MeV}}
{m_s(m_c)+m_d(m_c)}\right]^2$, see \cite{Buras:2014sba}.}
, is defined
as $F(x)=P_X X_0(x)+P_Y Y_0(x)+P_Z Z_0(x)+P_E E_0(x)$ with $P_i=r_i^{(0)}+r_i^{(6)}R_6
+r_i^{(8)}R_8$ ($i=0,X,Y,Z,E$).
We also update numerical values for %short distance
the coefficients $r_i^{(0)}$, $r_i^{(6)}$ and $r_i^{(8)}$,
%We adopt their numerical values
for $\alpha_s(M_Z)=0.1185$~\cite{PDG14} given in
Table~1 of Ref.~\cite{Buras:2014sba},
by reversing the sign of $r_0^{(j)}$ as done in Ref.~\cite{Hou:2005yb}.
We take %nonperturbative parameter %$R_8$ %that enters $\varepsilon^\prime/\varepsilon$,
%
%\begin{align}
$R_8 = 0.7$
% \label{eq:R8val}
%\end{align}
%
%obtained
from lattice~\cite{Blum:2012uk},
with the translation by Ref.~\cite{Buras:2014sba}.
There is still no reliable result from lattice QCD for $R_6$,
so we treat~\cite{Hou:2005yb,BijnensPrades} it as a parameter.
That is, for each value of $R_6=1.0, 1.5, 2.0, 2.5$,
we require $\varepsilon^\prime/\varepsilon$ to agree within
$1\sigma$ experimental error~\cite{PDG14},
\begin{align}
&\frac{\varepsilon^\prime}{\varepsilon}
 \simeq {\rm Re}\left( \frac{\varepsilon^\prime}{\varepsilon}\right)
= (1.66\pm 0.23)\times 10^{-3}.
\end{align}
%for each fixed value of $R_6$.

%
%We confirmed other observables, such as $\Delta m_{B_s}/\Delta m_{B_d}$, $\Delta m_K$ do not give further constraints in the parameter space of our interest.

%
%$B\to K^{(*)}\mu^+\mu^-$, $D$-$\bar D$ mixing??

Comments on other potentially important observables are in order.
In contrast to $\Delta m_{B_{d,s}}$,
$\Delta m_K$ is polluted by Long-Distance (LD) effects.
We have checked that there is no significant change of
the SD part from the SM value and
$(\Delta m_K)_{\rm SD}$ is still below the measured value.
$D^0$-$\bar D^0$ mixing is also subject to LD effects.
We checked that the SD contribution to the mixing amplitude
$M_{12}^D$ from $b^\prime$ (with $m_{b^\prime}\sim m_{t^\prime}$)
could be enhanced up to 3 times the SM value
in the parameter space of our interest,
but it is still well below the measured value of
$\Delta m_D$.

$B\to K^{(*)}\mu^+\mu^-$ observables are subject to
precise measurements at the LHC and severely constrain NP effects.
%The 4G $t^\prime$ could affect these observables mainly through
%$C_9$ and $C_{10}$.
We checked that the 4G $t'$ effects on %the Wilson coefficient
$C_9$ is %small,
within 5\% of the SM value ($\sim 4.3$)
in our parameter space. %of our interest.
The 4G effects on $C_{10}$ can be
as large as unity in some part of the target parameter space. % we consider.
However, adopting the model independent constraint
in Ref.~\cite{Altmannshofer:2014rta}, we checked that the changes
are within $2\sigma$ for various modes.
It cannot explain $P_5^\prime$, nor $R_K$, anomalies.

\section{Results}

%Taking the measurement value of $|V_{ub}|$ extracted from
%exclusive $B$ decay modes, we could not find any
%``solution'' for New Physics for our purpose. Thus,
%we use $|V_{ub}^{\rm ave}| = 0.00415 \pm 0.00049$ throughout this work.

To illustrate the connection between $B_d\to\mu^+\mu^-$
and $K_L \to \pi^0\nu\bar\nu$, we explore two scenarios
(see Fig.~1):
%for $V_{t'd}^*V_{t'b} \equiv r_{db}e^{i\phi_{db}}$ (Eq.~(\ref{eq:rdbphidb})):
%
\begin{itemize}
\item Scenario S1: \ \ $r_{db}e^{i\phi_{db}} = 0.00040\, e^{i\,330^\circ}$\\
% This scenario considers
 $\mathcal B(B_d\to\mu^+\mu^-)\gtrsim 3\times$ SM,
 with $e^{i\phi_{db}}$ complex;
\item Scenario S2: \ \ $r_{db}e^{i\phi_{db}} = 0.00045\, e^{i\,260^\circ}$\\
 $\mathcal B(B_d\to\mu^+\mu^-) \sim$ SM,
 $\phi_{db}$ is near maximal CPV;
% allowing larger $K_L \to \pi^0\nu\bar\nu$.
%\item Scenario S2$^\prime$: \ \ $r_{db}e^{i\phi_{db}} = 0.00075\, e^{i\,260^\circ}$\\
% Extend S2 to larger $\mathcal B(B_d\to\mu^+\mu^-)$
% by larger $r_{db}$.
\end{itemize}

With formulas and data input given in Sec.~II,
we plot in Fig.~2[left] the region in the
$|V_{t'd}^* V_{t's}|$--$\arg(V_{t'd}^* V_{t's})$
or $r_{ds}$--$\phi_{ds}$ plane allowed by various constraints
for S1.
The golden-hued (very) light shaded regions are for 1(2)$\sigma$
of %Eq.~(\ref{eq:BsmumuLHC}),
the $B_s \to\mu^+\mu^-$ mode.
Other constraints, labeled by the process, cut in at certain regions:
 ${\cal B}(K_L \to \mu\mu)_{\rm SD}$
 at the upper-left corner, and just right of center;
 $\phi_s = -0.049(-0.088)$ at 1(2)$\sigma$ cuts off
  %the 1(2)$\sigma$-allowed $B_s \to\mu^+\mu^-$
 near center of right-hand side,
 and a tiny sliver in first quadrant.
% while the $K^+ \to \pi^+\nu\nu$ constraint at 90\% CL cuts off further
%  the 2$\sigma$-allowed $B_s \to\mu^+\mu^-$ in similar region.
%
%The $K^+ \to \pi^+\nu\nu$ constraint %does not enter, as it
%is weaker than all other constraints.
%
The remaining 1$\sigma$ contours for $B_s \to\mu^+\mu^-$
correspond to $3.5\times 10^{-9}$ (central-left region)
and $2.2\times 10^{-9}$ (4th quadrant extending into 1st quadrant) in rate,
and for 2$\sigma$ contours,
$4.3\times 10^{-9}$~\cite{B2mumu-LHCbCMS} from 1st to 2nd quadrant
and $1.6\times 10^{-9}$ in 4th quadrant only.
We find that $R = \Delta m_{B_d}/\Delta m_{B_s}$ does
not provide further constraint within 2$\sigma$.

The allowed region of Fig.~2[left] is further overlaid, in Fig.~2[right],
by the constraints of $\varepsilon_K$, $\varepsilon'/\varepsilon$,
and give $K_L \to \pi^0\nu\nu$ contours in red-solid, labeled
by BR values in $10^{-10}$ units.
Note that ``15'' %on the righthand side
is just above the nominal GN bound of Eq.~(\ref{eq:GNbound}),
while the region $\lesssim$ SM strength is marked by
red-dash lines %on lefthand side
with label ``SM''.
The $\varepsilon_K$ constraint, plotted in shaded blue with
 theoretical error (experimental error negligible),
prefers small $|V_{t'd}^* V_{t's}|$ values, except two ``chimneys''
where the phase of $V_{t'd}^* V_{t's}$ is small for one near $180^\circ$,
and the other is tilted in the fourth quadrant.
 %because $\phi_{db} \simeq 330^\circ$.
%
The $\varepsilon'/\varepsilon$ constraint is more subtle,
because of the less known~\cite{BijnensPrades} hadronic parameter $R_6$
(we fix $R_8 \simeq 0.7$~\cite{Blum:2012uk}). %Eq.~(\ref{eq:R8val})).
We illustrate~\cite{Hou:2005yb} with $R_6 = 1.0,\ 1.5,\ 2.0,\ 2.5$,
in ascending order of green bands %in Fig.~2[right]
determined by experimental error of $\varepsilon'/\varepsilon$.

\begin{figure*}[t!]
\centering
%\includegraphics[width=60mm,height=40mm]{plots/twidth.pdf}
%\vspace{30mm}
{\includegraphics[width=80mm]{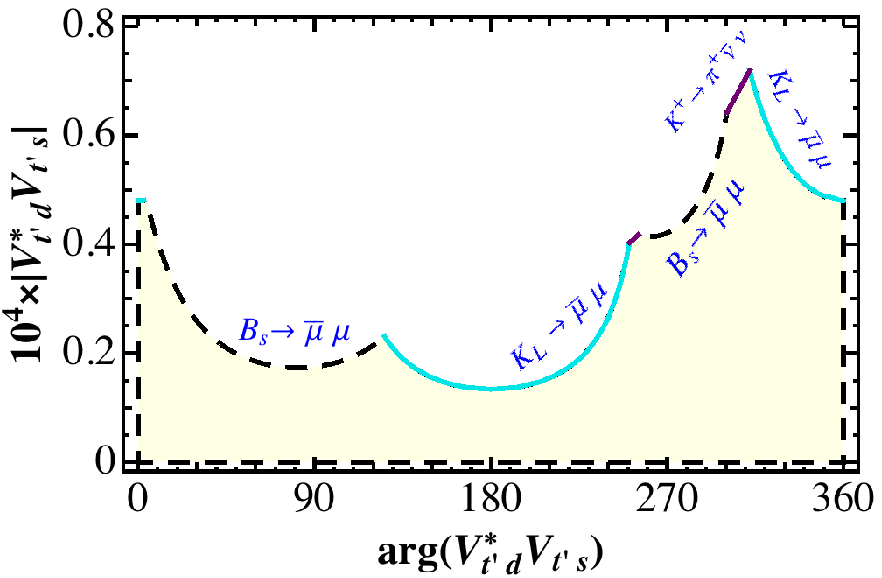}
 \includegraphics[width=80mm]{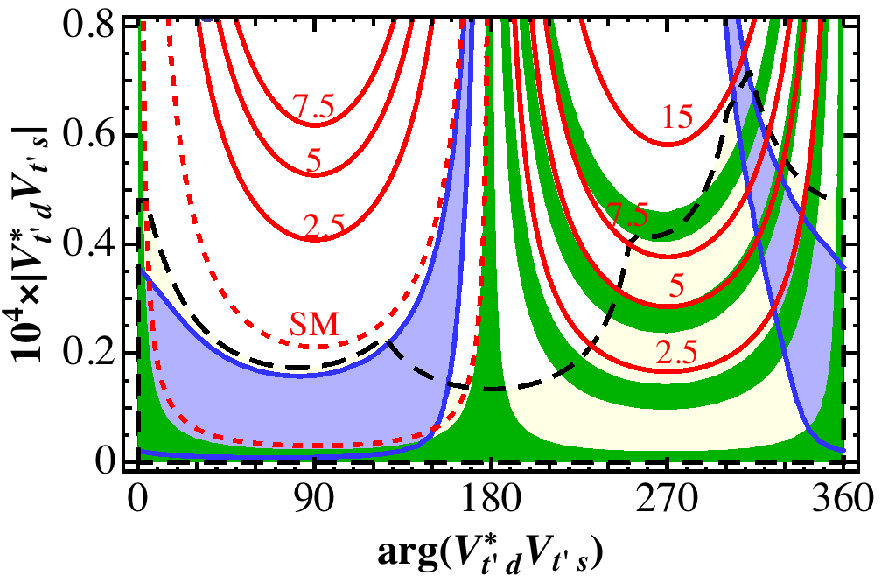}
}
\caption{
Scenario S2,
 $r_{db}\,e^{i\phi_{db}} = 0.00045\,e^{i\,260^\circ}$ (SM-like $B_d \to \mu^+\mu^-$):
 [left] Similar to Fig.~2a, where
%   skipping the intermediate step of Fig.~2a.
 for the 4th quadrant of interest, the $2\sigma$ dashed line is
 for $B_s \to \mu^+\mu^-$ and solid is for $K_L \to \mu^+\mu^-$
 (plus a bit from $K^+ \to \pi^+\nu\bar\nu$);
 [right] Similar to Fig.~2b, with $\varepsilon_K$ (blue-shaded)
 $\varepsilon^\prime/\varepsilon$ (green bands) and
 $K_L \to \pi^0\nu\bar\nu$ (red labelled contours) overlaid.
} \label{fig:fig3}
\end{figure*}

First, we observe that
 the $\varepsilon_K$ and $\varepsilon'/\varepsilon$ constraints
 disfavor the possible enhancements for $K_L \to \pi^0\nu\nu$
 when $\arg(V_{t'd}^* V_{t's})$ is in the first two quadrants.
Second, if one keeps all constraints to 1$\sigma$,
then $K_L \to \pi^0\nu\nu$ could reach a factor $\sim 7$ above SM,
with modest $R_6$ values.
% $K_L \to \pi^0\nu\nu$ would be close to SM expectation.
However, if one allows larger $R_6$ (up to 2.5)
as well as 2$\sigma$ variations,
the $\varepsilon_K$ ``chimney'' in the 4th quadrant allows
$K_L \to \pi^0\nu\nu$ to be enhanced up to 1/3, even 1/2, the GN bound.
%(the line labeled 15 ($\times 10^{-10}$)),
%with $R_6$ reaching 2 or 2.5 in strength.
There is a correlation between
larger $K_L \to \pi^0\nu\nu$ and smaller $B_s\to\mu\mu$.
If KOTO observes $K_L \to \pi^0\nu\nu$ shortly after
reaching below the GN bound, a rather large $R_6$ value
could be implied.
One argument for larger $K_L \to \pi^0\nu\nu$ or smaller $B_s\to\mu\mu$
is for larger values of $|V_{t'd}^* V_{t's}|$:
since $|V_{t'd}| \sim 0.005$, to have $|V_{t's}| > |V_{t'd}|$
would demand $|V_{t'd}^* V_{t's}| \gtrsim 0.25\times 10^{-4}$.

We see the %intricacies and
prowess, still, of the various
kaon measurements, with $K_L \to \pi^0\nu\nu$ as the main frontier
($K^+\to\pi^+\nu\bar\nu$ did not enter discussion),
on a par with the ongoing $B_{d,\,s} \to\mu\mu$ and $\phi_s$ measurement efforts.

For Scenario S2, where $B_d\to\mu\mu$ is taken as consistent with SM,
but $\phi_{db} \equiv \arg(V_{t'd}^*V_{t'b})\simeq 260^\circ$
is close to maximal CPV phase (in our convention, $V_{t'b}$ is real)
with $K_L \to \pi^0\nu\bar\nu$ in mind,
we plot in Fig.~3[left] the results corresponding to
Fig.~2[left].
The regions marked by long dashed lines and very lightly shaded
are all beyond 1$\sigma$ level, indicating more tension,
including in $R = \Delta m_{B_d}/\Delta m_{B_s}$.
The $B_s\to\mu\mu$ constraint at 2$\sigma$
is interspersed with the ${\cal B}(K_L \to \mu\mu)_{\rm SD}$ constraint,
plus short segments from $K^+\to \pi^+\nu\nu$.
As in Fig.~2[right],
we overlay the constraints of $\varepsilon_K$, $\varepsilon'/\varepsilon$,
as well as $K_L \to \pi^0\nu\nu$ contours, in Fig.~3[right].
%The 2$\sigma$ \emph{lower} $\phi_s = -0.04$ gives bound from below,
%while on the left it is bounded by 2$\sigma$ $B_s\to \mu^+\mu^-$
%($4.3 \times 10^{-9}$~\cite{B2mumu-LHCbCMS}) from above,
%and on the right, bounded from above by
%$B_s \to \mu^+\mu^-$ ($1.6 \times 10^{-9}$),
%a sliver of $K^+\to \pi^+\nu\nu$ (90\% CL),
%and $\phi_s = 0.015$ (1$\sigma$), respectively.
%
Again, $K_L \to \pi^0\nu\nu$ cannot get enhanceed in first two quadrants.
For the blue-shaded ``chimney'' in 4th quadrant,
as the $R_6$ value rises, $K_L \to \pi^0\nu\nu$
could get enhanced even up to GN bound,
but $B_s\to\mu\mu$ would become relatively suppressed,
and there is some tension with
SD contribution to $K_L \to \mu\mu$.
Note that ${\cal B}(B_d\to\mu\mu) \sim$ SM in this case.
Here, having $|V_{t's}| > |V_{t'd}|$
would demand $|V_{t'd}^* V_{t's}| \gtrsim 0.32\times 10^{-4}$,
hence in favor of larger $K_L \to \pi^0\nu\nu$.

We have marked a point S2$^\prime$ in Fig.~1,
which has same $\phi_{db} \simeq 260^\circ$ as S2,
but enhances $B_d\to\mu\mu$ by a larger
$r_{db} \equiv |V_{t'd}^*V_{t'b}| \simeq 0.00075$.
The trouble with S2$^\prime$ %, though allowed by Fig.~1,
is that $\Delta m_{B_d}/\Delta m_{B_s}$ ratio becomes inconsistent at 2$\sigma$ level,
which we do not consider as viable.
However, from S2 towards S2$^\prime$,
one could enhance $B_d\to\mu\mu$ while
$K_L \to \pi^0\nu\nu$ is more easily enhanced up to GN bound.
The cost would be some tension in $\Delta m_{B_d}/\Delta m_{B_s}$.
%We again skip the intermediate step corresponding to Fig.~2[left],
%and plot the results in Fig.~3[right]
%with 2$\sigma$ allowed region marked by long dashed lines:
%$K_L\to\mu\mu$ at left, right and center,
%$B_s\to\mu\mu$ in first quadrant,
%and $K^+\to \pi^+\nu\bar\nu$ in 3-to-4 quadrant.
%
%There is practically no solution in the first two quadrants,
%and we note that small $|V_{t'd}V_{t's}|$ and $R_6 \simeq 1$
%is not allowed in this scenario.
%
%At the cost of tension in $\Delta m_{B_s}/\Delta m_{B_d}$,
%$K_L \to \pi^0\nu\nu$ can reach the GN bound for large $R_6$ values,
%while $K^+\to \pi^+\nu\bar\nu$ would also be at E949 limits,
%which bodes well for both KOTO and NA62.
%
%Here, having $|V_{t's}| > |V_{t'd}|$
%would demand $10^4\times|V_{t'd}^* V_{t's}| \gtrsim 0.9$,
%hence $K_L \to \pi^0\nu\nu$ must be close to GN bound.

In all these discussions, $\phi_s$ is well within
range of the 3 fb$^{-1}$ result of LHCb, Eq.~(\ref{eq:phis_LHCb-3}).

\section{\label{sec:Discussion} \boldmath
Discussion and Conclusion\protect\\}

We are interested in the correlation between
$B_d\to \mu^+\mu^-$ and $K_L\to \pi^0\nu\bar\nu$ in 4G,
as constrained by $B_s \to \mu\mu$ and $\phi_s$.
Scenario S1 illustrates enhanced $B_d\to \mu^+\mu^-$
with generic $V_{t'd}^*V_{t'b}$.
Every measurement other than $B_d\to \mu^+\mu^-$
would be close to SM expectation,
and a mild enhancement of $K_L \to \pi^0\nu\nu$ is possible.
But it would take some while for KOTO to reach this sensitivity.
Larger $K_L \to \pi^0\nu\nu$ correlates with smaller $B_s\to\mu^+\mu^-$,
as well as larger hadronic parameter $R_6$.
The $\phi_s$ constraint basically
suppresses the phase of $V_{t's}^*V_{t'b}$.

It could happen that $B_d\to \mu^+\mu^-$ ends up SM-like,
which is illustrated by Scenario S2.
In the 4G framework that accounts (within 1$\sigma$)
for the $\sin2\beta/\phi_1$ ``anomaly'', this occurs when
$\phi_{db} \equiv \arg(V_{t'd}^*V_{t'b})$ phase is near maximal,
which is of interest for enhancing $K_L \to \pi^0\nu\nu$,
a purely CPV process.
We find that $K_L \to \pi^0\nu\nu$ can be enhanced up
to practically the GN bound at the cost of large $R_6$,
while staying within the $\phi_s$ constraint.
There is the same correlation of larger $K_L \to \pi^0\nu\nu$
for smaller $B_s\to \mu^+\mu^-$.
While the S2$^\prime$ point would push
$\Delta m_{B_d}/\Delta m_{B_s}$ beyond 2$\sigma$ tolerance,
some $|V_{t'd}^*V_{t'b}| \equiv r_{db}$ value below 0.00075
could still enhance $B_d\to \mu^+\mu^-$ a bit from SM,
but $K_L \to \pi^0\nu\nu$ can more easily
saturate the Grossman-Nir bound,
% (which may be preferred if one demands $|V_{t's}| > |V_{t'd}|$),
with implication that $K^+\to \pi^+\nu\bar\nu$ is towards the
large side allowed by E949,
$B_s \to \mu\mu$ is visibly suppressed,
while $R_6$ must be sizable.
This would clearly be a bonanza situation for faster discovery!

We have used 4G for illustration~\cite{foot},
since it supplies $V_{t's}$ and $V_{t'd}$ that affect
$b\to s$ and $b\to d$ transitions, and induces correlations with
$s\to d$ transitions.
It is generally viewed that the fourth generation is ruled out
by the SM-like Higgs boson production cross section.
But we have argued~\cite{Hou:2013btm} that the
Higgs boson does not enter the low energy processes
discussed here, hence these processes are independent
\emph{flavor} checks.
Furthermore, loopholes exist for the SM-Higgs interpretation~\cite{MHK}.

%The correlations and possible tensions in the 4G framework is
%taken as an illustration of the possible correlation between
%$b\to s$, $b\to d$, and $s\to d$ processes.
If one does not accept 4G, we stress that
$B_d\to \mu^+\mu^-$ may well turn out have enhanced rate.
Whatever new flavor physics one resorts to,
there is the myriad of constraints of Sec.~II.
We believe no model can survive intact and ``without blemish''~\cite{Mimura}.
Thus, the modes $B_{d,\,s}\to \mu^+\mu^-$, %$B_s\to \mu^+\mu^-$,
$\phi_s$ and $K_L \to \pi^0\nu\nu$ provide ``pressure tests"
to our understanding of flavor and $CP$ violation,
%
%Maybe all measurements become consistent with SM in 10 years, but
where genuine surprises may emerge.
Though differences must exist,
we believe there would be correlations between the
above four modes in any New Physics model
with a limited set of new parameters.
The NA62 experiment has started~\cite{NA62} running.
If $K^+ \to \pi^+\nu\nu$ turns out to be above the 90\% CL limit from E949.
the GN bound for $K_L \to \pi^0\nu\nu$ moves up,
making things more interesting for KOTO,
where the aim~\cite{KOTO} for the 2015 run is to reach
the GN bound around $1.4\times 10^{-9}$.
%This is a volatile situation that ought to be watched.

In conclusion,
enhanced $B_d^0 \to \mu^+\mu^-$
could correlate with enhanced $K_L \to \pi^0\nu\bar\nu$
up to the Grossman-Nir bound in the 4th generation model.
$B_s^0 \to \mu^+\mu^-$ becomes somewhat suppressed,
with CPV phase $\phi_s \simeq 0$.
Together with $K^+\to\pi^+\nu\bar\nu$,
these measurements would  provide ``pressure tests"
to our understanding of flavor and $CP$ violation
for any New Physics model.
They should be followed earnestly in parallel to the
scrutiny of the nature of the 125 GeV boson at LHC Run~2.

\vskip0.3cm
\noindent{\bf Acknowledgement}.  WSH is supported by the the
Academic Summit grant NSC 103-2745-M-002-001-ASP of the
National Science Council, as well as by grant NTU-EPR-103R8915.
MK is supported under NTU-ERP-102R7701 and NSC 102-2112-M-033-007-MY3.
FX is supported under NSC 102-2811-M-002-205, as well as by NSFC under
grant No. 11405074.
We thank T. Yamanaka for discussions that stimulated this work.

\vskip0.1cm
\noindent{\bf Note Added}.  During the revision,  the long-waited result of
$B_6^{(1/2)}$ appeared \cite{Buras:2015yba}, extracted  from a new 
lattice calculation %of $A_0$ 
carried out by RBC-UKQCD collaboration \cite{Bai:2015nea}, which 
indicates a small $R_6$.

%%%%%

%%%%%


\begin{thebibliography}{99}
\bibitem{foot}
The $t'$ mass is large, but always come with
CKM factors, hence remains perturbative.
The loop functions may be modified by the strong
Yukawa coupling, but our purpose to
illustrate the power of CKM4 should remain valid.
\end{thebibliography}
\end{document}